\documentclass{PoS}

\title{Physics results from dynamical overlap fermion simulations}

\ShortTitle{Physics results from dynamical overlap fermion simulations}

\author{\speaker{Shoji Hashimoto}\\
        High Energy Accelerator Research Organization (KEK), 
        Tsukuba 305-0801.\\
        E-mail: \email{shoji.hashimoto@kek.jp}}

\abstract{
  I summarize the physics results obtained from large-scale dynamical overlap
  fermion simulations by the JLQCD and TWQCD collaborations.
  The numerical simulations are performed at a fixed global topological
  sector; the physics results in the $\theta$-vacuum is reconstructed by
  correcting the finite volume effect, for which the measurement of the
  topological susceptibility is crucial.
  Physics applications we studied so far include a calculation of chiral
  condensate, pion mass, decay constant, form factors, as well as (vector and
  axial-vector) vacuum polarization functions and nucleon sigma term.
}

\FullConference{The XXVI International Symposium on Lattice Field Theory\\
		 July 14-19 2008\\
		 Williamsburg, Virginia, USA}

\begin{document}

\section{Introduction}
Many important aspects of the low energy dynamics of QCD are related
to the chiral symmetry and its spontaneous breaking in the QCD vacuum.
Lattice QCD is a promising tool to actually {\it simulate} the low
energy regime of QCD and to reproduce the spontaneous symmetry
breaking and related phenomena, provided that the chiral symmetry is
preserved on the lattice.
This was the prime motivation for the JLQCD and TWQCD collaborations to
embark on the QCD simulations with exact chiral symmetry, {\it i.e.}
dynamical overlap fermion simulations, despite the enormous
computational costs they require. 

Theoretically, with the dynamical overlap fermions, one may
investigate the relation between chiral symmetry breaking and 
low-lying modes of the Dirac operator, as suggested by the Bank-Casher
relation and the chiral random matrix theory.
Also, the quantities related to the topology of the SU(3) gauge field,
such as the topological susceptibility, may be studied using fermionic
proves using the singlet axial-Ward-Takahashi identity.

Phenomenologically, the chiral symmetry plays an important role 
in the chiral extrapolation of lattice data for many physical
quantities.
Since the use of the {\it continuum} chiral perturbation theory is
justified, the chiral extrapolation is largely simplified when the
chiral symmetry is exactly preserved on the lattice.
In this talk, I will discuss a few examples that demonstrate the
chiral extrapolation performed using the next-to-next-to-leading order 
chiral perturbation theory formulae.
We have other new applications for which the exact chiral symmetry
plays a crucial role.
They include a calculation of the $VV-AA$ vacuum polarization functions
and an extraction of the strange quark content from the (partially
quenched) nucleon masses.

This project was started in 2006 when the present supercomputer
system, that comprises Hitachi SR11000 and IBM Blue Gene/L, was
installed at KEK. 
A report of the project at an earlier stage was presented by Hideo
Matsufuru at Lattice 2007 \cite{Matsufuru:2007uc}, which mainly
described the simulation itself.
The present talk discusses about the physics applications from
the project in some detail.
For individual contributions, see their own write-ups
\cite{Chiu:2008kt,Matsufuru_lat08,Noaki:2008gx,Ohki:2008ge,Shintani_lat08,Kaneko_lat08,Yamada_lat08}.

\section{Simulation strategy and status}
The overlap-Dirac operator $D(0)\equiv\rho/a[1+X/\sqrt{X^\dagger X}]$
(with the Wilson kernel $X\equiv aD_W-\rho$) 
\cite{Neuberger:1997fp,Neuberger:1998wv}
provides a theoretically ideal fermion formulation on the lattice, as
it has an exact chiral symmetry at finite lattice spacings 
\cite{Luscher:1998pq} while satisfying the index theorem.
Implementation of the overlap operator is however non-trivial, since
the definition contains a discontinuity at the zero of the denominator
$X^{\dagger}X$ (or the zero of the hermitian Wilson-Dirac operator
$H_W\equiv\gamma_5 X$).
One usually approximates the sign function $\mathrm{sgn}(H_W)$ using
polynomial or rational functions, but the near-zero eigenmodes must be
specially treated. 
At finite lattice spacings, it is known that there are finite
density of near-zero modes of $H_W$ \cite{Edwards:1998sh}, that
means that the computational cost to identify the near-zero modes
grows as quickly as $O(V^2)$ and prevents one from performing large scale
simulations using the overlap fermion.

The near-zero modes are associated with a dislocation of the background
gauge field and thus are unphysical (an explicit example is found in
\cite{Berruto:2000fx}).
We suppress them by introducing extra heavy Wilson fermions
(and associated ghosts) \cite{Vranas:2006zk,Fukaya:2006vs}.
They produce a factor $\det[H_W^2/(H_W^2+\mu^2)]$ to the Boltzmann
factor ($\mu$ is a mass of the associated ghosts) and naturally
suppress the near-zero modes.
The numerical cost for the overlap fermion simulation is then
dramatically reduced especially when the lattice volume is large.

Since the topology change (or the change of the number of zero-modes
of the overlap-Dirac operator) may occur only when a small eigenvalue
of $H_W$ changes its sign, the suppression factor strictly prohibits
the tunneling among different topological sectors, as far as the Monte
Carlo algorithm is based on a continuous change of the gauge field
configuration. 
This is in fact a property of the continuum QCD, {\it i.e.} topology
change cannot occur through continuous deformation of the gauge field.

Dynamical overlap fermions have been attempted by several authors
\cite{Fodor:2003bh,DeGrand:2004nq,Cundy:2005pi}, in which the treatment
of the topology tunneling is one of the main issues.
Among them, our simulations have reached the level to produce broad
physics results for the first time.
On a lattice of size $\sim (\mbox{2~fm})^3$ we simulate dynamical
fermions with several quark masses between $m_s/6$ and $m_s$ with
$m_s$ the physical strange quark mass.
In addition to the runs for two-flavor QCD, which is already published 
\cite{Aoki:2008tq}, we have completed a series of runs for 2+1-flavor QCD.
The main reason that made this possible is the topology-fixing, as
well as the computational power provided by the BlueGene/L. 

The two-flavor runs were done at $\beta=2.30$ using the Iwasaki gauge
action on a $16^3\times 32$ lattice.
The lattice spacing determined through the Sommer scale $r_0$ =
0.49~fm is 0.118(2)~fm.
For each of six sea quark masses covering the region $m_s/6\sim m_s$, we
accumulated 10,000 HMC trajectories in the trivial topological sector
$Q=0$.
(For a check of the fixed topology effect, we also produced
configurations of $Q=-2$ and $-4$ at one sea quark mass. 
For details, see below.)

The 2+1-flavor runs were carried out at the same $\beta$ value on a
$16^3\times 48$ lattice. 
This corresponds to the lattice spacing $a$ = 0.108(2)~fm.
We took two strange quark masses.
For each of them, five up and down quark masses are taken covering the
similar quark mass region.
In total, ten runs of each length 2,500 HMC trajectories have been
accumulated at $Q=0$.

\section{Topology issues}
If the topological charge is frozen in the QCD simulation, is there
any problem?
The answer to this question is obviously ``yes''.
The QCD vacuum is required to be the $\theta$-vacuum, a superposition
of different topological sectors, in order to satisfy the cluster
decomposition property.
Therefore, we cannot sample the correct QCD vacuum unless the
topological charge fluctuates sufficiently.
This is a serious problem for everyone doing dynamical QCD simulation
aiming at approaching the continuum limit, since the topological
charge would not change near the continuum limit irrespective of which
fermion formulation one employs.

We believe that the solution (one of the solutions, at least) to this
problem is to reconstruct the $\theta$-vacuum physics from the fixed
topology simulations \cite{Aoki:2007ka}.
This is based on the observation that the effect of fixed topological
charge is a finite volume effect of $O(1/V)$,
which can be derived from a simple Fourier transform between the 
$\theta$ vacuum and the fixed $Q$ ``vacuum''.
For instance, the partition function at a given topological charge $Q$
is obtained using a saddle point expansion as
\begin{equation}
  Z_Q = \frac{1}{2\pi\chi_tV}\exp\left[-\frac{Q^2}{2\chi_tV}\right]
  \left[
    1-\frac{c_4}{8V\chi_t}
    +O\left(\frac{1}{(\chi_tV)^2},\frac{Q^2}{(\chi_tV)^2}\right)
  \right],
\end{equation}
where $\chi_t$ is the topological susceptibility and $c_4$
characterizes a deviation from the Gaussian distribution at a finite
volume. 
Therefore, if one can calculate $\chi_t$ and $c_4$ (and higher order
terms if necessary), the $\theta$-vacuum can be reconstructed by simply
adding the different topological sectors as
$Z(\theta)=\sum_Q Z_Qe^{-i\theta Q}$.
The method to calculate these key parameters is discussed shortly.

The relation between the $\theta$ and $Q$ vacuums can be extended to
any physical quantities.
For example, for a (CP-even) two-point function, one can express it in a given 
topological charge $Q$ in terms of its counterpart in the $\theta$(=0) vacuum 
$G(0)$ and its second and fourth derivatives $G^{(2)}(0)$ and $G^{(4)}(0)$ as
\cite{Aoki:2007ka,Brower:2003yx}
\begin{equation}
  \label{eq:G_Q}
  G_Q=G(0)+G^{(2)}(0)\frac{1}{2\chi_tV}
  \left[1-\frac{Q^2}{\chi_tV}-\frac{c_4}{2\chi_t^2V}\right]
  +G^{(4)}(0)\frac{1}{8\chi_t^2V^2}+\cdots
\end{equation}
up to higher order corrections in $1/V$.
Therefore, provided that the $\theta$-dependence of the quantity of interest is
known, the $\theta$-vacuum physics can be reconstructed.
The $\theta$-dependence can be obtained for the quantities that can be
analysed using chiral perturbation theory; if it is not applicable one needs
the data at several $Q$ to extract $G^{(2)}(0)$ for instance.

\begin{figure}[tb]
  \centering
  \includegraphics*[width=10cm]{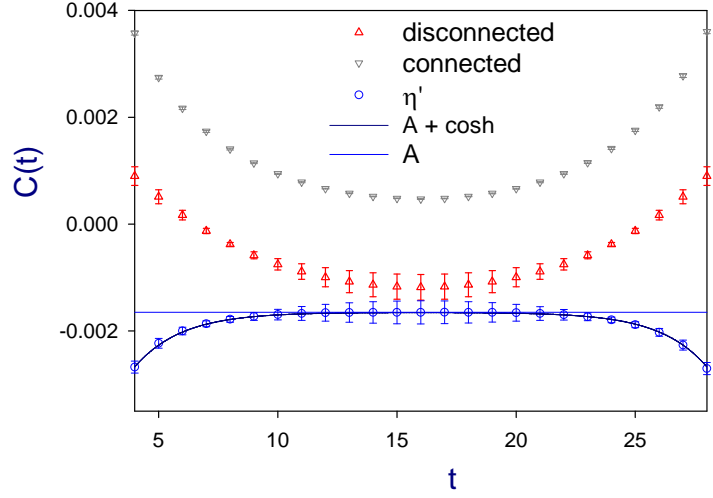}
  \caption{
    Pseudo-scalar density correlator (\protect\ref{eq:chi_t}) summed over
    spatial separation and plotted as a function of $t$.
    The data for $N_f=2$ QCD at $am=0.025$.
    Solid line shows a fit with a constant giving the topological
    susceptibility. 
  }
  \label{fig:mPmP}
\end{figure}

By applying the formula (\ref{eq:G_Q}) for a topological charge density
correlator, or equivalently a correlator of flavor-singlet pseudo-scalar
densities $mP(x)$, one obtains a relation 
\begin{equation}
  \label{eq:chi_t}
  \lim_{x\to\infty}\langle mP(x)mP(0)\rangle_Q =
  -\frac{1}{V}\left(\chi_t-\frac{Q^2}{V}+\cdots\right)+O(e^{-m_{\eta'}x}),
\end{equation}
where $\langle\cdots\rangle_Q$ stands for an expectation value in the given
$Q$. 
The topological susceptibility $\chi_t$ can be extracted from the asymptotic
value of the correlator in (\ref{eq:chi_t}).
One expects a negative constant when $Q=0$, which is intuitively understood as
follows. 
When one finds a positive topological charge density at the origin, it is more
likely that a negative value is observed at a far distant point $x$, if the
net topology is fixed to $Q=0$. 
The value will be suppressed when the volume is increased.

An example obtained in two-flavor QCD is shown in Figure~\ref{fig:mPmP}
\cite{Aoki:2007pw}.
We observe a clear plateau for all six different ensembles with different sea
quark masses.
At $am=0.050$, we obtain a consistent results from the configurations with
different topological charges $Q=0$, $-2$, and $-4$.

\begin{figure}[tb]
  \centering
  \includegraphics*[width=10cm]{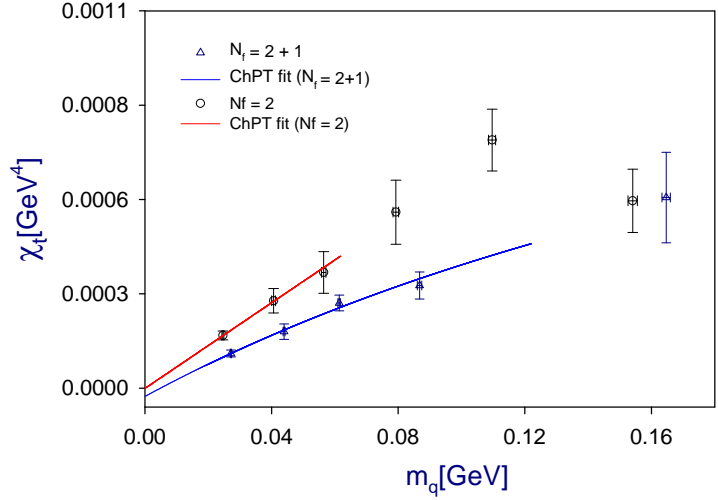}
  \caption{
    Topological susceptibility as a function of sea quark mass.
    The data for $N_f=2$ (circles) and $N_f=2+1$ (triangles) are shown.
    The solid curves represent a fit with the form of chiral effective theory.
  }
  \label{fig:chi_t}
\end{figure}

Sea quark mass dependence of $\chi_t$ is expected to behave as
$\chi_t=\Sigma/(1/m_u+1/m_d+1/m_s)$ at the leading order of chiral
perturbation theory \cite{Leutwyler:1992yt}.
In two-flavor QCD with degenerate sea quark mass $m$, it becomes
$\chi_t=m\Sigma/2$.
Figure~\ref{fig:chi_t} shows the results for 2- and 2+1-flavor QCD
\cite{Chiu:2008kt}. 
The data show a clear linear dependence on the sea quark mass.
The fits to two- or 2+1-flavor formula yield the value of chiral condensate
$\Sigma$.
After a renormalization, we obtain 
$\Sigma(\mathrm{2~GeV})$ = [242(5)(10)~MeV]$^3$ ($N_f$=2) and 
$\Sigma(\mathrm{2~GeV})$ = [240(5)(2)~MeV]$^3$ ($N_f$=2+1).

\section{Physics applications}
\subsection{Chiral condensate}
For the lattice calculation of the chiral condensate, the exact chiral
symmetry plays a crucial role.
Without the exact symmetry the scalar density operator on the lattice
$(\bar{\psi}\psi)^{lat}$ may mix with the identity operator under radiative
corrections. 
This leads to a power (cubic) divergence in the matching onto the
corresponding continuum operator $(\bar{\psi}\psi)^{cont}$, which prevents one
from calculating this quantity on the lattice with any useful precision.

\begin{figure}[tb]
  \centering
  \includegraphics*[width=9cm]{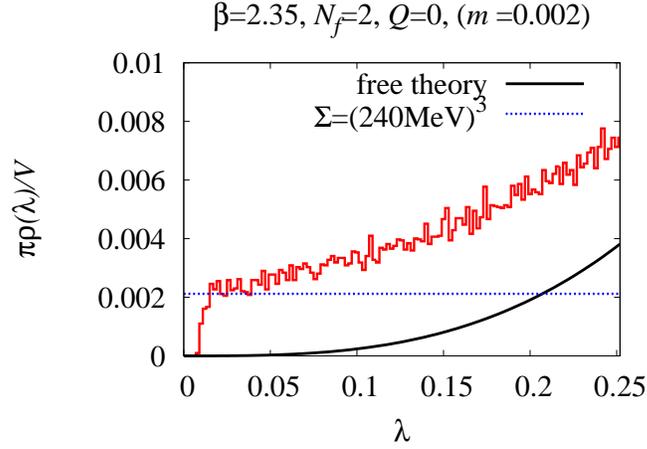}
  \caption{
    Spectral density $\rho(\lambda)$ as a function of $\lambda$,
    calculated from lattice data in the $\epsilon$-regime.
    Banks-Casher relation infers a constant at $\lambda=0$, which is
    not of course satisfied at a finite volume, but the remnant is
    clearly seen.
    The horizontal dashed line is an expectation for a nominal value
    of $\Sigma$.
    }
  \label{fig:spectrum}
\end{figure}

Thanks to the exact chiral symmetry, we are able to precisely calculate this
quantity from several different sources.
They contain the use of the Banks-Casher relation (as demonstrated in
Figure~\ref{fig:spectrum}), the spectrum of low-lying
eigenvalues (with a help of the chiral random matrix theory), the meson
correlators in the $\epsilon$-regime, the topological susceptibility, and the
Gell-Mann-Oaks-Renner (GMOR) relation.

In the $\epsilon$-regime, where pion Compton wavelength is longer than the
spatial extent of the lattice, the dynamical degrees of freedom is
dominated by the zero-momentum modes of pions.
In this circumstance, the low-lying eigenmode spectrum can be obtained using
the Chiral Random Matrix Theory (ChRMT).
The chiral condensate can be extracted by matching the spectrum calculated on
the lattice with the ChRMT prediction.
We obtain 
$\Sigma^{\overline{\mathrm{MS}}}(\mathrm{2~GeV})$ =
[251(7)(11)~MeV]$^3$ \cite{Fukaya:2007fb,Fukaya:2007yv}.
The meson correlator calculated in the $\epsilon$-regime can also be used to
determine the chiral condensate $\Sigma$ as well as the pion decay constant
$F$. 
We obtain 
$\Sigma^{\overline{\mathrm{MS}}}(\mathrm{2~GeV})$ =
[240(4)(7)~MeV]$^3$
and 
$F$ = 87(6)(8)~MeV \cite{Fukaya:2007pn}.

\begin{figure}[tb]
  \centering
  \includegraphics*[width=8cm]{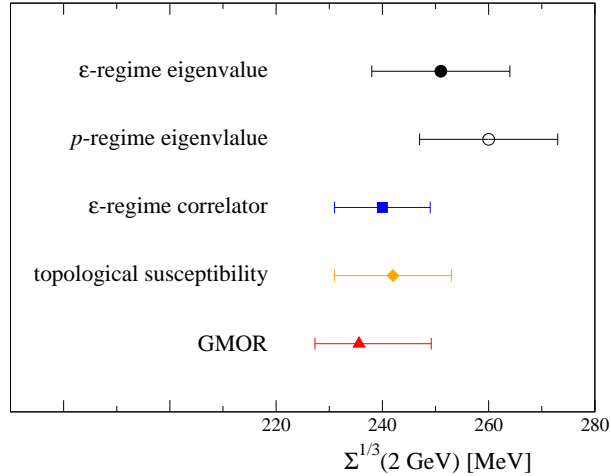}
  \caption{
    Chiral condensate calculated in two-flavor QCD with several different
    methods: 
    $\epsilon$-regime eigenvalue spectrum \cite{Fukaya:2007fb,Fukaya:2007yv},
    $p$-regime eigenvalue spectrum \cite{Fukaya:2007yv},
    $\epsilon$-regime meson correlator \cite{Fukaya:2007pn},
    topological susceptibility \cite{Chiu:2008kt},
    GMOR relation \cite{Noaki:2008iy}.
  }
  \label{fig:condensate}
\end{figure}

Figure~\ref{fig:condensate} compares the chiral condensate extracted with
various observables in two-flavor QCD.
They are in good agreement.
This strongly indicates that the low-energy dynamics of QCD is indeed
determined by the pion degrees of freedom and therefore that the
chiral symmetry is spontaneously broken, which is usually assumed when
constructing the chiral effective theory. 

\subsection{$m_\pi$ and $f_\pi$}
Near the massless limit of quarks the chiral effective theory is
expected to provide a good description of the low-energy QCD.
For larger pion masses, one must calculate loop corrections and also introduce
irrelevant operators to subtract associated divergences.
This leads to the chiral perturbation theory (ChPT), which is organized as an
expansion in terms of small $m_\pi^2$ and $p^2$.
The region of the convergence of this chiral expansion is not known a priori.
Using lattice QCD, one can test the expansion and identify the region of
convergence. 
With the exact chiral symmetry, the test is conceptually cleaner, since no
additional term to describe the violation of chiral symmetry has to be
introduced.
(With other fermion formulations, this is not the case. The unknown correction
terms are often simply ignored.)

For the pion mass $m_\pi$ and decay constant $f_\pi$ the expansion is given as
\begin{eqnarray}
  \label{eq:chiral_exp_mpi}
  \frac{m_\pi^2}{m_q} & = & 2B
  \left[ 1 + \frac{1}{2} x\ln x + c_3 x + O(x^2) \right],
  \\
  \label{eq:chiral_exp_fpi}
  f_\pi & = & f
  \left[ 1 - x\ln x + c_4 x + O(x^2) \right],
\end{eqnarray}
where $m_\pi$ and $f_\pi$ denote the quantities after the corrections while 
$m$ and $f$ are them at the leading order.
The expansions (\ref{eq:chiral_exp_mpi}) and (\ref{eq:chiral_exp_fpi}) may be
written in terms of either
$x\equiv 2m^2/(4\pi f)^2$, 
$\hat{x}\equiv 2m_\pi^2/(4\pi f)^2$, or
$\xi \equiv 2m_\pi^2/(4\pi f_\pi)^2$
(we use a notation of $f_\pi$ = 131~MeV).
Theoretically, they all give an equivalent description at this order,
while the convergence behavior may depend on the expansion parameter.

\begin{figure}[tb]
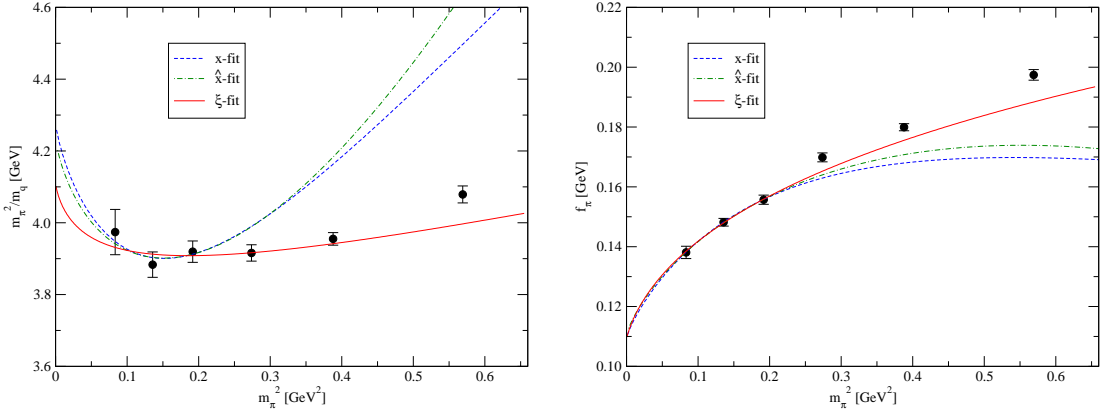

  \centering
  \includegraphics*[width=7cm]{figure/mp2r_nlo.eps}
  \hspace*{4mm}
  \includegraphics*[width=7cm]{figure/fps_nlo.eps}
  \caption{
    Comparison of chiral expansion in terms of $x$, $\hat{x}$ and
    $\xi$.
    The plots represent $m_\pi^2/m_q$ (left) and $f_\pi$ (right).
    Fits of the three lightest data points with the NLO ChPT formulae
    (\protect\ref{eq:chiral_exp_mpi}) and (\protect\ref{eq:chiral_exp_fpi})
    are shown. 
  } 
  \label{fig:mpifpi}
\end{figure}

Figure~\ref{fig:mpifpi} shows the comparison of different expansion
parameters \cite{Noaki:2008iy}.
The fit curves are obtained by fitting the three lightest data points with the
three different fit parameters.
They all provide equally precise description of the data in the region of the
fit.
If we look at the heavier quark mass region, however, it is clear that only
the $\xi$-expansion gives a reasonable function and others miss the data
points largely.
This clearly demonstrates that at least for these quantities the convergence
of the chiral expansion is much improved by the use of the $\xi$-parameter
than the conventional choice, {\it i.e.} the $x$-expansion.

\begin{figure}[tb]
  \centering
  \includegraphics*[width=7.5cm]{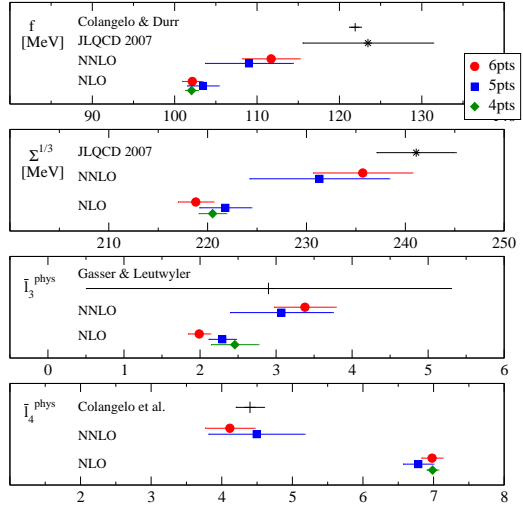}
  \caption{
    Low energy constants of two-flavor QCD as determined from the (N)NLO
    analysis of our lattice data.
    The results are obtained using 4, 5, or 6 lightest sea quark mass data.
    Phenomenological numbers are taken from 
    Colangelo-Durr \cite{Colangelo:2003hf},
    Colangelo {\it et al.} \cite{Colangelo:2001df};
    the previous JLQCD results were obtained from the meson correlators in the
    $\epsilon$-regime \cite{Fukaya:2007pn}.
  } 
  \label{fig:LEC}
\end{figure}

Therefore, we consider only the $\xi$-expansion when we extend the analysis to
the next-to-next-to-leading order (NNLO).
I do not write down the NNLO terms here, but an important point is that the
coefficient of the {\it leading log} terms of the form $(\xi\ln\xi)^2$ is
determined solely from ChPT and thus is not a fit parameter.
The terms of the form $\xi^2\ln\xi$ and $\xi^2$ have to be determined by the
fit. 
It is indeed possible by a combined fit of $m_\pi^2/m_q$ and $f_\pi$;
we obtain the low energy constants as shown in Figure~\ref{fig:LEC}.

It is clear from Figure~\ref{fig:LEC} that the results depend on the
order of the chiral expansion, {\it i.e.} NLO or NNLO, especially for
the NLO low energy constants $\bar\ell_3$ and $\bar\ell_4$.
This is because the NNLO term appears with a relatively large
numerical coefficient, such as
$\bar\ell_3 \xi(1-9/2\xi\ln\xi)$, for instance.
A large effect is then expected for the determination of $\bar\ell_i$
unless $\xi$ is much less than 0.1.
The NNLO terms are mandatory in the analysis of lattice data.

We are currently extending the NNLO analysis to the partially quenched data in
two-flavor QCD and to the 2+1-flavor QCD data \cite{Noaki:2008gx}.

\subsection{Pion form factor}
Pion vector and scalar form factors provide another simple testing ground for
the chiral dynamics of pion.
They are defined as
\begin{eqnarray}
  \langle\pi(p')|V_\mu|\pi(p)\rangle & = & (p_\mu+p'_\mu)F_V(q^2),\\
  \langle\pi(p')|S|\pi(p)\rangle & = & F_S(q^2), \quad q_\mu\equiv p'_\mu-p_\mu,
\end{eqnarray}
for the vector current $V_\mu$ and the scalar density $S$.
At small momentum transfer $q^2$, the form factors are characterized by the
charge and scalar radii, $\langle r^2\rangle_V^\pi$, and
$\langle r^2\rangle_S^\pi$, as
\begin{eqnarray}
  F_V(q^2) & = & 1+\frac{1}{6}\langle r^2\rangle_V^\pi q^2 + O(q^4), \\  
  F_S(q^2) & = & F_S(0)
  \left[ 1+\frac{1}{6}\langle r^2\rangle_S^\pi q^2 + O(q^4) \right].
\end{eqnarray}
Note that the vector form factor $F_V(q^2)$ is normalized at $q^2=0$ due to
the current conservation.

We use the {\it all-to-all} quark propagator technique \cite{Foley:2005ac} to
calculate the form factors \cite{Kaneko:2007nf,Kaneko_lat08}. 
The main advantage of this technique is, of course, that one can calculate
the disconnected diagram contribution to the scalar form factor.
Furthermore, it substantially reduces the statistical error in the calculation
through an average over the source point.

\begin{figure}[tb]
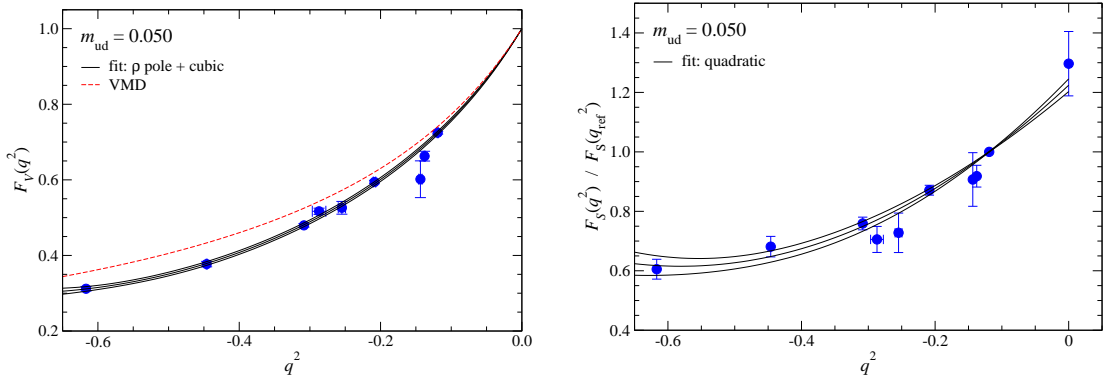

  \centering
  \includegraphics*[width=7cm]{figure/pff_v_vs_q2_m0050.eps}
  \hspace*{4mm}
  \includegraphics*[width=7cm]{figure/pff_s_vs_q2_m0050.eps}
  \caption{
    Pion vector (left) and scalar (right)form factors calculated at
    $am=0.050$.
  } 
  \label{fig:pff}
\end{figure}

We calculate the form factors at four lightest sea quark mass configurations
available in two-flavor QCD.
The data at a fixed quark mass are shown in Figure~\ref{fig:pff}.
For the vector form factor we attempt a fit ansatz motivated by the pole
dominance
$F_V(q^2)=1/(1-q^2/m_V^2)+c_1 q^2+\cdots$
with $m_V$ the vector meson mass obtained at the same quark mass.
This is justified by the analyticity of the form factor and also by a
reasonable assumption that the higher pole contributions are well
approximated by a Taylor series. 
The data are, in fact, nicely fitted with this function as shown in
Figure~\ref{fig:pff} (left).
The higher pole effects are small but visible.
For the scalar form factor there is no obvious pole, and we use a simple
polynomial of $q^2$ to fit the data (right panel).

\begin{figure}[tb]
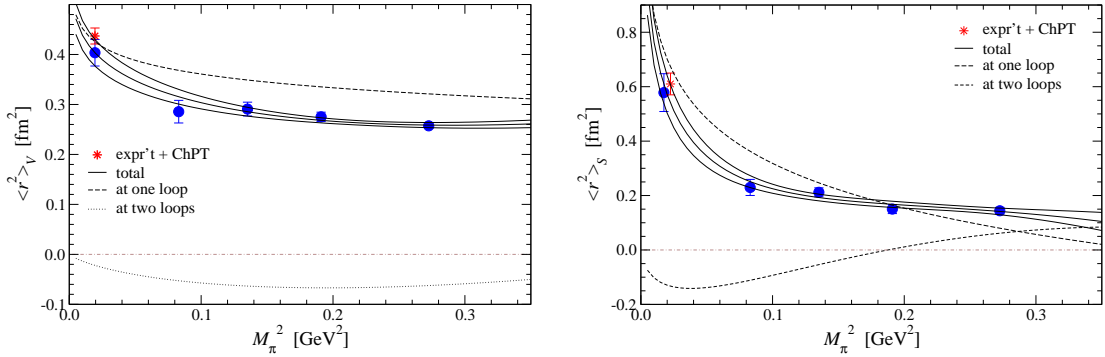

  \centering
  \includegraphics*[width=7cm]{figure/r2_v_vs_Mpi2.meas-pole+cubic.r2+c_v_nnlo.eps}
  \hspace*{4mm}
  \includegraphics*[width=7cm]{figure/r2_s_vs_Mpi2.meas-pole+cubic.r2_v_s_c_v_nnlo.eps}
  \caption{
    Vector (left) and scalar (right) radii of pion as a function of $M_\pi^2$.
    Solid curves show the fits with the NNLO ChPT formulae.
    Their breakdown into NLO (dashed) and NNLO (short dashed) contributions
    are also shown.
  } 
  \label{fig:radii}
\end{figure}

The charge and scalar radii thus extracted are plotted as a function
of $m_\pi^2$ in Figure~\ref{fig:radii} together with the fits with the
NNLO ChPT formulae. 
For these quantities, the ChPT predicts a logarithmic divergence in the chiral
limit $\sim (1/f_\pi^2)\ln m_\pi^2$ with a fixed numerical coefficient.
The data do not show the expected divergent behavior, but the fit is possible
and gives a value consistent with the phenomenological value at the point of
physical pion mass.
This suggests that the divergent behavior shows up in the even smaller quark
mass region.
We note that the fit is possible only when we include the NNLO terms,
which contain additional logarithmic divergences in the chiral limit.

\subsection{Vacuum polarization functions}
The vacuum polarization functions defined through
\begin{equation}
  \int d^4x e^{iqx}\langle 0|J_\mu(x)J_\nu^\dagger(0)|0\rangle
  =
  (g_\mu q^2-q_\mu q_\nu) \Pi_J^{(1)}(Q^2) - q_\mu q_\nu \Pi_J^{(0)}(Q^2)
\end{equation}
contain rich information of QCD dynamics in both perturbative and
non-perturbative regimes.
($J_\mu$ is either a vector $V_\mu$ or an axial-vector $A_\mu$ current.
We restrict ourselves in the flavor non-singlet currents, though the flavor
indices are suppressed in the expressions.)
Experimentally, the vacuum polarization function can be extracted from
$e^+e^-$ annihilation or from $\tau$ decay processes.

The spontaneous chiral symmetry breaking can be probed with the vacuum
polarization functions using the Weinberg sum rules \cite{Weinberg:1967kj},
such as 
\begin{eqnarray}
  f_\pi^2 & = & -\lim_{Q^2\to 0} Q^2
  \left[\Pi_V^{(1+0)}(Q^2)-\Pi_A^{(1+0)}(Q^2)\right], \\
  S & = & -\lim_{Q^2\to 0} \frac{\partial}{\partial Q^2} Q^2
  \left[\Pi_V^{(1+0)}(Q^2)-\Pi_A^{(1+0)}(Q^2)\right], 
\end{eqnarray}
where $S$ stands for the Peskin-Takeuchi's parameter used in the context of
the precision electroweak measurement \cite{Peskin:1990zt}.
In chiral effective theory, it is related to a low energy constant $L_{10}$.
These sum rules can be obtained in the limit of massless quarks.
Because of their $VV-AA$ structure, they vanish when the chiral symmetry is not
spontaneously broken.

Along these lines, another interesting quantity is the electromagnetic mass
difference of pion, which is expressed in this limit as
\begin{equation}
  \label{eq:DGMLY}
  \Delta m_\pi^2 = - \frac{3\alpha_{EM}}{4\pi f_\pi^2}
  \int_0^\infty dQ^2 Q^2 \left[\Pi_V^{(1+0)}(Q^2)-\Pi_A^{(1+0)}(Q^2)\right],
\end{equation}
which is known as the 
Das-Guralnik-Mathur-Low-Young sum rule \cite{Das:1967it}.

For a lattice calculation of these quantities, exact chiral symmetry is
essential, since we have to extract a tiny difference between the vector and
axial-vector channels.
In the calculation using the overlap fermions, we checked that even the
lattice artifacts precisely cancel between $V$ and $A$ \cite{Shintani:2008qe}.

\begin{figure}[tb]
  \centering
  \includegraphics*[width=8cm]{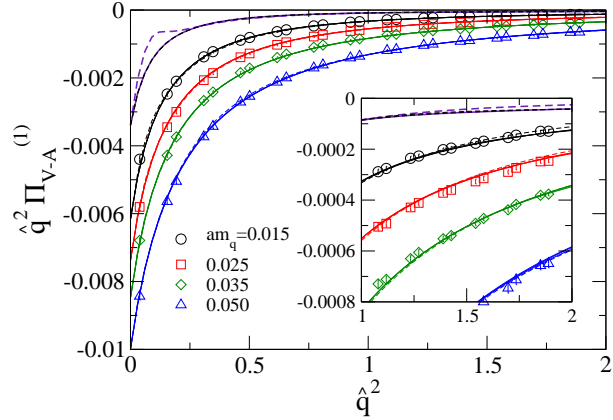}
  \caption{
    Difference of the vacuum polarization functions
    $\Pi_V^{(1)}(Q^2)-\Pi_A^{(1)}(Q^2)$ multiplied by $Q^2$.
    Data are shown for different quark masses; the chiral limit is also
    plotted by a solid (linear extrapolation) and a dashed (chiral logs)
    curve. 
  }
  \label{fig:v-a}
\end{figure}

In Figure~\ref{fig:v-a} we plot the results for the difference
$\Pi_V^{(1)}(Q^2)-\Pi_A^{(1)}(Q^2)$.
The chiral limit is taken by assuming a linear quark mass dependence (solid
curve) or by using the one-loop ChPT formula (dashed curve).
The use of ChPT is slightly questionable, because the lowest (non-zero) value
of $Q^2$ on our lattice is already as large as (320~MeV)$^2$ and the second
lowest $\sim$(650~MeV)$^2$ is clearly out of the range.
The high $Q^2$ region, on the other hand, may be analyzed using the operator
product expansion (OPE).
OPE is also used to extend the region of $Q^2$ to infinity as required in
(\ref{eq:DGMLY}). 
For the pion mass difference, we finally obtain
$m_\pi^2$ = 992(12)($^{+\ \ 0}_{-135}$)(149)~MeV$^2$
(see \cite{Shintani:2008qe} for details),
which may be compared with the experimental value 1,261~MeV$^2$.

Another use of the vacuum polarization function is an extraction of the strong
coupling constant by matching the lattice data with the OPE expression in the
perturbative regime.
For this purpose we consider the Adler function
$D_J(Q^2)\equiv -Q^2 d\Pi_J(Q^2)/dQ^2$, 
which is free from ultraviolet divergence, and thus renormalization scheme
independent. 
It means that one can use the expressions obtained in the continuum
perturbation theory to describe the lattice data.
(In the practical analysis, we directly use $\Pi_J(Q^2)$ instead, to avoid
numerical derivative in terms of $Q^2$.
We then float a divergent constant as a fit parameter.)

In the lattice calculation, it is necessary to subtract lattice artifacts
mainly due to non-conserved lattice currents.
Non-perturbative subtraction is possible by utilizing several different
momentum configurations giving the same $Q^2$ \cite{Shintani:2008ga}.

\begin{figure}[tb]
  \centering
  \includegraphics*[width=8cm]{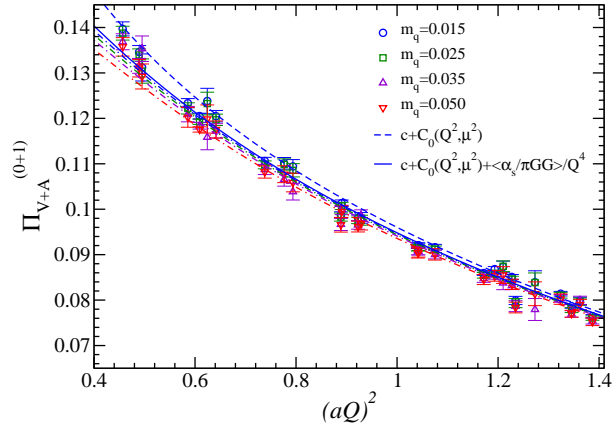}
  \caption{
    Vacuum polarization function $\Pi_{V+A}^{(1)}(Q^2)$ as a function of
    $Q^2$.
    The lattice data at different quark masses are shown together with the fit
    curve with the OPE formula (solid curve).
  }
  \label{fig:v+a}
\end{figure}

Lattice results are shown in Figure~\ref{fig:v+a}.
The data in high $Q^2$ regime are fitted with the perturbative expression
supplemented by OPE
\begin{equation}
  \Pi_J^{\mathrm{pert}}(Q^2) = c + C_0(Q^2,\mu^2) + \frac{C_m^J(Q^2)}{Q^2}
  + C_{\bar{q}q}^J(Q^2) \frac{\langle m\bar{q}q\rangle}{Q^4}
  + C_{GG}(Q^2) \frac{\langle(\alpha_s/\pi)GG\rangle}{Q^4} + \cdots,
\end{equation}
where perturbative expansion of $C_0$, $C_m$, {\it etc} is known to
$\alpha_s^2$ in the $\overline{\mathrm{MS}}$ scheme.
$c$ is the divergent constant term.

In the analysis, we may use the value of the chiral condensate
$\langle\bar{q}q\rangle$ obtained through other quantities as described above.
We then obtain
$\Lambda_{\overline{\mathrm{MS}}}^{(2)}$ = 234(9)($^{+16}_{-\ 0}$)~MeV 
in two-flavor QCD \cite{Shintani:2008ga}.
This result is compatible with previous lattice calculations in
two-flavor QCD,
250(16)(16)~MeV \cite{DellaMorte:2004bc} or
249(16)(25)~MeV \cite{Gockeler:2005rv}.

\subsection{Nucleon structure}
The last application of our dynamical overlap fermion simulations reported in
this talk is a calculation of the nucleon sigma term
\begin{equation}
  \label{eq:sigma}
  \sigma_{\pi N}=m_{ud}\langle N|\bar{u}u+\bar{d}d|N\rangle
\end{equation}
and the strange quark content
\begin{equation}
  \label{eq:y}
  y = \frac{2\langle N|\bar{s}s|N\rangle}{
    \langle N|\bar{u}u+\bar{d}d|N\rangle}.
\end{equation}
For $\sigma_{\pi N}$ both connected and disconnected quark lines contribute,
while only the disconnected diagram contributes to the strange quark content.

Instead of directly calculating the matrix elements in (\ref{eq:sigma}) and
(\ref{eq:y}), we utilize the Feynman-Hellman theorem.
Namely, we calculate the derivatives of nucleon mass in terms of valence or
sea quark mass.
They are related to the connected and disconnected contributions as
\begin{eqnarray}
  \frac{\partial M_N}{\partial m_{\mathrm{val}}} & = &
  \langle N|\bar{u}u+\bar{d}d|N\rangle_{\mathrm{conn}}, \\
  \frac{\partial M_N}{\partial m_{\mathrm{sea}}} & = &
  \langle N|\bar{u}u+\bar{d}d|N\rangle_{\mathrm{disc}}
  ( \simeq 2 \langle N|\bar{s}s|N\rangle ),
\end{eqnarray}
where the subscripts ``conn'' and ``disc'' stand for the connected and
disconnected contributions to the matrix element, respectively.
The disconnected piece becomes identical to the strange matrix element
$2\langle N|\bar{s}s|N\rangle$ when the sea quark mass is set equal to the
strange quark mass.
To obtain the derivatives we need partially quenched 
($m_{\mathrm{val}}\ne m_{\mathrm{sea}}$) data set.
For the fit of the quark mass dependence, we use the partially quenched chiral
perturbation theory calculated at one-loop for the nucleon mass
\cite{Chen:2001yi,Beane:2002vq}.

\begin{figure}[tb]
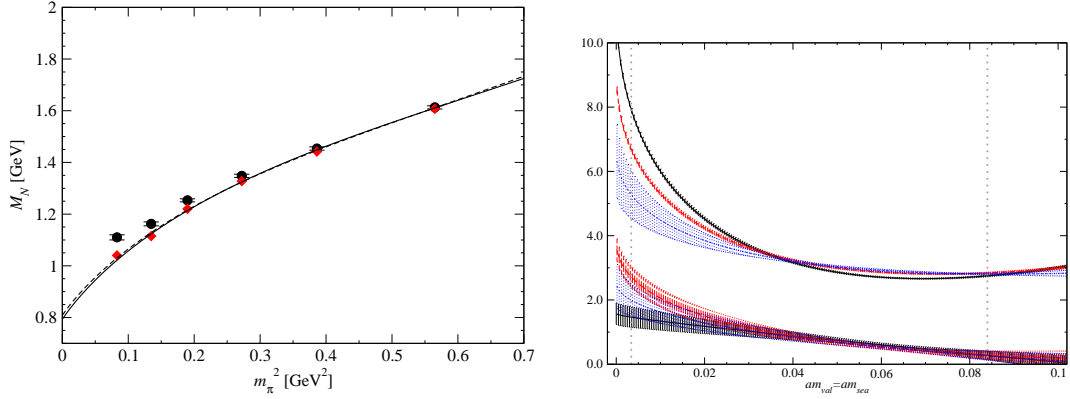

  \centering
  \includegraphics*[width=7cm]{figure/FSE_FIT0.eps}
  \hspace*{4mm}
  \includegraphics*[width=6.5cm]{figure/Fig7.eps}
  \caption{
    (Left) Nucleon mass in two-flavor QCD.
    The data after the finite volume correction are shown by red symbols.
    The curves represent the fit with baryon ChPT formulae.
    (Right) The connected (upper) and disconnected (lower) contributions to
    the sigma term as a function of the quark mass.
    Sea and valence quarks are taken to be equal.
    Physical up/down and strange quark masses are shown by vertical lines.
  } 
  \label{fig:sigma}
\end{figure}

Figure~\ref{fig:sigma} (left panel) shows the nucleon mass calculated on our
two-flavor QCD configurations as a function of $m_\pi^2$ \cite{Ohki:2008ff}.
Finite volume effect is corrected using the expectation from ChPT 
\cite{Ali Khan:2003cu}. 
The data are nicely fitted with the ChPT formula at $O(p^3)$ or at $O(p^4)$, 
through which we obtain 
$\sigma_{\pi N}$ = 52(2)($^{+20}_{-\ 2}$)($^{+5}_{-0}$)~MeV
(for further details, see \cite{Ohki:2008ff}).

The derivatives can be obtained after fitting the partially quenched data set
with the corresponding ChPT formulae.
The results are shown in Figure~\ref{fig:sigma} (right panel).
The plot clearly shows that the disconnected contribution is smaller than the
connected contribution.
In particular, the disconnected piece is tiny for the mass corresponding to
the strange quark.
As a result the $y$ parameter we obtained is small:
$y$ = 0.030(16)($^{+6}_{-8}$)($^{+1}_{-2}$).

Previous lattice calculations gave rather large values for $y$:
0.66(15) \cite{Fukugita:1994ba}, 0.36(3) \cite{Dong:1995ec},
0.59(13) \cite{Gusken:1998wy}.
The problem in these calculations is that the critical mass depend
significantly on the sea quark mass when one uses the Wilson-type fermion due
to the lack of chiral symmetry.
One can misidentify this unphysical effect with the physical sigma term.
(We note that the problem persists even when one directly calculates the matrix
element rather than using the Feynman-Hellman theorem.
That is because the scalar density operator made up with valence quarks may
mix with that with sea quarks.
One has to subtract this operator mixing to obtain the correct results
\cite{Ohki:2008ge}.)

After subtracting this effect, UKQCD obtained a value consistent with zero,
$y=-0.28(33)$, albeit a large error due to the subtraction
\cite{Michael:2001bv}.
Our setup is free from this problem because of the use of the overlap fermion
for both valence and sea quarks.
The exact chiral symmetry plays a key role here, too.

\section{Conclusions}
In this talk I demonstrate that the dynamical overlap fermion may provide a
clean approach to the problems related to the chiral symmetry and its
spontaneous breaking.
Despite the large numerical cost it requires, large scale simulation is
feasible with the computational resources of $O$(10~TFlops).
So far, we have completed runs on a 16$^3\times$48 and started test runs on a
larger volume, 24$^3\times$48.

The key to this success is the new strategy of treating the global topological
charge. {\it i.e.} we fix the topology during the simulation and reproduce the
$\theta$-vacuum physics by correcting an induced finite volume effect.
In doing so, the key observation is that the topological susceptibility can be
correctly extracted from simulations at a fixed topology.

The problem of topology freezing will become common among all lattice
simulations as one sufficiently approaches the continuum limit, simply because
this is the property of the continuum QCD.
Our work is a first attempt to overcome that situation.

The exact chiral symmetry plays a crucial role in the calculation of many
important physical quantities.
For instance, we have calculated the chiral condensate using several
methods for the first time and confirmed the consistency among them, which
provides the most fundamental test of the chiral effective theory.
We are extending the test to various quantities, such as pion mass, decay
constant, charge radius and so on.
(A calculation of the kaon $B$ parameter is already published
\cite{Aoki:2008ss}.) 
There are other interesting applications for which the exact chiral symmetry
is required; so far we studied the $VV-AA$ vacuum polarization and the nucleon
sigma term.
We plan to further pursue in this direction.

The author is supported in part by Grant-in-aid for Scientific
Research (No.~18340075).

\end{document}